\documentclass[reqno]{article}%12pt%
\usepackage{alltt}
\usepackage{graphicx}
\title{Towards the demystification of quantum interference}
\author{Peter Leifer}
\date{Cathedra of Informatics, Crimea State Engineering and
Pedagogical University, \\
21 Sevastopolskaya st., 95015 Simferopol, Crimea, Ukraine; \\
leifer@bezeqint.net }
\begin{document}
\maketitle
\begin{abstract}
It has been shown that velocity of propagation of wave front cannot coincide with observable velocity of quantum particles.
 It is additional argument leads to conclusion that phase wave of de Broglie cannot be associated with single ``elementary" particle like electron. Therefore quantum interference under linear superposition cannot describe energy distribution in extended quantum particles. Essentially new approach is required in order to establish non-linear relativistic wave equations with soliton-like solutions.
\end{abstract}
\vskip 0.1cm
\section{Introduction}
It is very symptomatic that last time arose many articles discussing fundamental and ``elementary" physical notions interesting in view of deep crisis in quantum physics \cite{Schroer}. Schroer writes in fact about crisis in particle physics but the reason is rooted much deeper. I would like to attract attention to some articles
\cite{Valentini,Huang,Nassif,Kryukov1,Kryukov2} where given attempts to discover
some new relations between relativity and quantum theory. My approach to the unification \cite{Le1,Le2,Le3,Le4,Le5} is more close to ``functional relativity" of \cite{Kryukov1,Kryukov2} under some specification and reservations.

Here I would like
to give additional argument that quantum amplitudes of some event in fixed setup could not be used by itself for derivation correct quantum dynamics and morphogenesis of extended
quantum particles. Rather \emph{velocities of deformation of these amplitudes} arose under infinitesimal variations of setup parameters lead to new relativistic quantum dynamics
\cite{Le5}.

De Broglie wave concept
of particles \cite{dB} and Schr\"odinger's equation for hydrogen atom paved \cite{Sch1} the way to corpuscular-wave duality of matter. Quantum theory solved a lot of fundamental problems, but
(as it happen with fundamental theory), it posed number of deeper questions.
Even first steps in wave picture of quantum particles brought sudden surprises.
First of all there was a big discrepancy between intuitive Schr\"odinger's imagination
and real properties of ``corpuscular waves".
Initially Schr\"odinger thought that there is a possibility to build stable wave packet
from plane waves of de Broglie that may be treated as the wave model of localized
electron; he understood soon that it is impossible. Only in some special
case of quantized harmonic oscillator he could build such stable wave packet moving
(in average) like material point under Hooke's elastic force \cite{Sch2}.

The impossibility to get wave description of localizable particle led to
probabilistic interpretation of the wave function. In fact, this is  the fork point
changing all character of fundamental physics: state vector is treated as amplitudes of
probability for particle to be in some particular state. This paradigm is the
source of all fundamental unsolved problems mentioned above: measurement problem,
localization, divergences, etc. However, applications of quantum theory are so
convincible and prolific that any attempts to find answers on ``children questions"
are frequently treated as a pathology. Nevertheless, we should analyze these problems
again and again till all situation will be absolutely clear without any references to
the beauty of quantum mystic \cite{Feynman1}.

\section{Wave front moves towards particle!}
Let me ask: where the wave nature of electrons in the original Davisson-Germer
experiments is emerged? An electron gun emits accelerated electrons moving inside
the gun and between the gun and crystal without any wave properties like classical
charged particles. This may be confirmed by the classical calculations of momentum, etc., in old oscilloscope or TV set tubes and electron lamps. Only after scattering on the crystal one sees the character wave interference pattern according to Bragg's law of the scattering
with de Broglie relation $\lambda = h/p$. Therefore if one has alternative coherent
sources of quantum particles (like in two-slits experiment or in multiple scattering
on lattice), there is evidence of wave behavior of \emph{quantum ensemble}.
We may conclude that beautiful de Broglie analysis of phase waves \cite{dB} is correct
and applicable to phase waves generated by periodic structure of crystal lattice
but it can not be attributed to electron itself. Why? The main reason of the
de Broglie's decision to identify particle with wave packet is that the coincidence of
group velocity with the velocity of ``body" \cite{dB}. This conclusion is based on
the analogy with Rayleigh's formula for the group velocity for the motion of the
wave packet in dispersive media
\begin{equation}
U= \frac{d \nu }{d(\frac{ \nu}{V })},
\end{equation}
where $\nu$ is frequency and $V$ is phase velocity. There are, however, in fact two reasons
why the de Broglie's waves can not be used for electron construction. One of the
reason is well known: linear wave packet is unstable and, therefore it is impossible
to identify it with stable particle like electron. This fact lead de Broglie to the
theory of the pilot-wave. There is the second reason why the plane waves
of de Broglie should not be literally refer to the state vector of
the electron itself but rather to covector (1-form) realized, say,
by a periodic crystal lattice \cite{G}.
The nice agreement of de Broglie prediction with experiments may be explained
that for the condition of diffraction it is important only the modulus of
relative velocity of electron and the lattice. The fact that velocity of
the wave front propagation $p_k=\frac{\partial W}{\partial q^k}$
and the electron velocity
$v^k=\frac{dq^k}{dt}$ compose sometimes the angle $\phi > \pi/2$ may be simply
proved.

Schr\"odinger pointed out that the
Hamilton-Jacobi equation for the material point with the unit mass
$m=1$
\begin{equation}
\frac{\partial W}{\partial t}+T(q^k,\frac{\partial W}{\partial
q^k})+V(q^k)=0
\end{equation}
may be understood as describing the surfaces of constant phase of
the action waves \cite{Sch1}. If
\begin{equation}
 W=-Et+S(q^k),
 \end{equation}
then one gets
\begin{equation}
T(q^k,\frac{\partial W}{\partial q^k})=E-V(q^k).
\end{equation}
The Hertz metric
\begin{equation}
dS^2=2 T(q^k,\frac{dq^k}{dt})dt^2=2(E-V(q^k))dt^2=g_{ik}dq^i dq^k.
\end{equation}
may be interpreted as a some effective ``index of refraction"
\begin{equation}
n= \frac{dS}{dt} = \sqrt{2 T} = \sqrt{g_{ik}v^i v^k}=\sqrt{2(h \nu -V)}
\end{equation}
for the action waves with the positive phase velocity
\begin{equation}
u=\frac{E}{\sqrt{2 T}} =\frac{E}{ \sqrt{g_{ik}v^i v^k}}=\frac{h
\nu}{\sqrt{2(h\nu -V)}}.
\end{equation}
Negative scalar product
\begin{equation}
p_k v^k=\frac{\partial W}{\partial
q^k}\frac{dq^k}{dt}=\frac{\partial W}{\partial t}= -E =-h\nu <0
\end{equation}
shows that the angle $\phi$ between momentum and the particle velocity should
obey inequality  $\phi > \pi/2$. This formal proof may be elementary confirmed
by the geometry of scattering on the periodic lattice.

I used above the original Schr\"odinger's approach and literal denotations
applicable in non-relativistic case. The proof in the relativistic case may be based on the
Schr\"odinger's method given in the book \cite{SchrB}.

Schr\"odinger discusses the conditions of approximate wave description
of material points moving along geodesic of curved space-time. The relativistic
analog of Hamilton-Jacobi equation for massive particle may be written as follows:
\begin{equation}
g_{ik}\frac{\partial I}{\partial x_i}\frac{\partial I}{\partial x_k}=
\frac{\partial I}{\partial x^k}\frac{\partial I}{\partial x_k}=1.
\end{equation}
The equation for particles governed by space-time curvature and by action
functional reads as follows
\begin{equation}
\frac{\partial I}{\partial x^k}=g_{ik}\frac{dx^i}{d\tau}, or \quad
\frac{dx^i}{d\tau}=g^{ik}\frac{\partial I}{\partial x^k}.
\end{equation}
In order to find its approximate wave solutions
Schr\"odinger had started with generalized wave equation
\begin{equation}
(g^{ik}\psi_{,k})_{,l}+\mu^2 \psi \sqrt{-g}=0
\end{equation}
taken in linear approximation. Namely, he put $\psi = A e^{i\phi}$ and took into
account the following conditions:

1. Relative variations of $g^{ik}$ and $A$ are small in comparison with the phase $\phi$
variations in any direction and so-called ``quantum potential" are negligible.

2. The phase $\phi$ is approximately a linear function  of coordinates, i.e.
$\phi=\sum_1^4 K_i x^i + K_5$ where $(K_i, K_5)$ are slow functions of $x^i$.

Then
\begin{equation}
\phi_i=\frac{\partial \phi}{\partial x^i}=K_i=(\omega, 2\pi \frac{cos\alpha}{\lambda},
2\pi \frac{cos\beta}{\lambda},2\pi \frac{cos\gamma}{\lambda})
\end{equation}
and, therefore, taking into account the signature of $g^{ik}$ it is easy to see that
\begin{equation}
\frac{dx^i}{d\tau}=g^{im}K_m=(\omega, -2\pi \frac{cos\alpha}{\lambda},
-2\pi \frac{cos\beta}{\lambda},-2\pi \frac{cos\gamma}{\lambda}).
\end{equation}
It means that there are at least two reasons why de Broglie-Schr\"odinger phase waves
correspond to multiple scattering processes and could not be attributed to single
quantum particle:
linear wave packet is unstable and generally the direction of wave front propagation
does not coincide with the direction of particle velocity. How we may restore quantum properties of single quantum particle which lurked behind the ensemble behaviour in
quantum interference?

The fundamental result of quantum interference phenomenon is that the variation of the quantum setup leads to deformation of resulting amplitude of some event. I try to establish the invariant law connecting \emph{velocities of state variation in Hilbert state space} with energy distribution expressed by relativistic non-linear equations whose solutions may be associated with single quantum particle \cite{Le1,Le2,Le3,Le4,Le5}.

\section{Conclusion}
It is shown that de Broglie-Schr\"odinger phase waves could not by itself describe energy distribution in single quantum particle. Only velocities of variations of these waves
may be connected with structure of quantum particles. The main technical problem
is to find non-Abelian gauge field arising from conservation law of the local Hailtonian
vector field. The last one may be expressed as parallel transport of local
Hamiltonian in projective Hilbert space $CP(N-1)$. Co-movable local ``Lorentz frame"
being attached to GCS is used for qubit encoding the result of comparison of the
parallel transported local Hamiltonian in infinitesimally close points. This
leads to quasi-linear relativistic field equations with soliton-like solutions
for ``field shell" in emerged dynamical space-time.


\vskip 0.2cm
\begin{thebibliography}{99}
\bibitem{Schroer}
B. Schroer, 0805.1911v2 [hep-th].
\bibitem{Valentini}
A. Valentini, arXiv:0811.0810v2 [quant-ph].
\bibitem{Huang}
Y.-S. Huang, arXiv:0901.3001v1 [physics.gen-ph].
\bibitem{Nassif}
C. Nassif, arXiv:0805.1201v3 [gr-qc].
\bibitem{Kryukov1}
A.A. Kryukov, Found. Phys. {\bf 36}, 175 (2006).
\bibitem{Kryukov2}
A.A. Kryukov, Found. Phys. {\bf 34}, 1225 (2004).
\bibitem{Le1}
P. Leifer, Found. Phys. {\bf 27}, (2) 261 (1997).
\bibitem{Le2}
P. Leifer, Annales de la Fondation Louis de Broglie, {\bf 32}, (1) 25
(2007).
\bibitem{Le3}
P. Leifer, Found.Phys.Lett., {\bf 18}, (2) 195 (2005).
\bibitem{Le4}
P. Leifer, JETP Letters, {\bf 80}, (5) 367 (2004).
\bibitem{Le5}
P. Leifer, arXiv:0901.2694v1 [physics.gen-ph].
\bibitem{dB}
L. de Broglie, Recherches sur la Th\'eorie des Quanta, (Ann. de Phys.
$10^e$ s\'erie, t.III (Janvier-F\'evrier 1925). Translated by A.F. Kracklauer,
\copyright AFK, 2004.
\bibitem{Sch1}
E. Schr\"odinger, Ann. Phys. {\bf 79}, 361, 1926.
\bibitem{Sch2}
E. Schr\"odinger, Nanurwissenschaften, {\bf 14}, H 28, 664, 1926.
\bibitem{Feynman1}
R.P. Feynman, \emph{QED The strange theory of light and matter}, Princeton,
New Jersey: Princeton University Press (1985).
\bibitem{G}
C.W. Misner, K.S. Thorne, J.A. Wheeler,{\it Gravitation}, W.H.Freeman
and Company, San Francisco, 1973.
\bibitem{SchrB}
E. Schr\"odinger, {\it Expanding Universes}, Cambridge at the University Press 1956.
\end{thebibliography}
\end{document}